\documentclass[twocolumn,superscriptaddress,floatfix,preprintnumbers,amssymb ,amsmath]{revtex4-1}
\usepackage{graphicx,amssymb}
\usepackage{color,ulem}\usepackage{bm}  
\usepackage{float}
\usepackage{siunitx}

\begin{document}
\title{Electron-phonon coupling of epigraphene at millikelvin temperatures}

\author{Bayan Karimi}
\affiliation{Pico group, QTF Centre of Excellence, Department of Applied Physics, Aalto University School of Science, P.O. Box 13500, 00076 Aalto, Finland}

\author{Hans He}
\affiliation{Department of Microtechnology and Nanoscience, Chalmers University of Technology, 412 96 Gothenburg, Sweden}

\author{Yu-Cheng Chang}
\affiliation{Pico group, QTF Centre of Excellence, Department of Applied Physics, Aalto University School of Science, P.O. Box 13500, 00076 Aalto, Finland}

\author{Libin Wang}
\affiliation{Pico group, QTF Centre of Excellence, Department of Applied Physics, Aalto University School of Science, P.O. Box 13500, 00076 Aalto, Finland}

\author{Jukka P. Pekola}
\affiliation{Pico group, QTF Centre of Excellence, Department of Applied Physics, Aalto University School of Science, P.O. Box 13500, 00076 Aalto, Finland}

\affiliation{Moscow Institute of Physics and Technology, 141700 Dolgoprudny, Russia}

\author{Rositsa Yakimova}
\affiliation{Department of Physics, Chemistry and Biology, Linkoping University, 581 83 Linköping, Sweden}

\author{Naveen Shetty}
\affiliation{Department of Microtechnology and Nanoscience, Chalmers University of Technology, 412 96 Gothenburg, Sweden}

\author{Joonas T. Peltonen}
\affiliation{Pico group, QTF Centre of Excellence, Department of Applied Physics, Aalto University School of Science, P.O. Box 13500, 00076 Aalto, Finland}

\author{Samuel Lara-Avila}
\affiliation{Department of Microtechnology and Nanoscience, Chalmers University of Technology, 412 96 Gothenburg, Sweden}

\author{Sergey Kubatkin}
\affiliation{Department of Microtechnology and Nanoscience, Chalmers University of Technology, 412 96 Gothenburg, Sweden}


\begin{abstract}
We investigate the basic charge and heat transport properties of charge neutral epigraphene at sub-kelvin temperatures, demonstrating nearly logarithmic dependence of electrical conductivity over more than two decades in temperature. Using graphene's sheet conductance as in-situ thermometer, we present a measurement of electron-phonon heat transport at mK temperatures and show that it obeys the $T^4$ dependence characteristic for clean two-dimensional conductor. Based on our measurement we predict the noise-equivalent power of $\sim 10^{-22}~{\rm W}/\sqrt{{\rm Hz}}$ of epigraphene bolometer at the low end of achievable temperatures.
\end{abstract}


\maketitle
Epitaxial graphene on SiC substrate (epigraphene) is an attractive scalable~\cite{Emtsev, Virojanadara} technology for high-quality graphene electronics~\cite{Tzalenchuk, Lara-Avila1, He1}. 
Using a recently reported doping technique ~\cite{He2}, epigraphene doped close to the Dirac point has shown to have great potential for astronomy-oriented terahertz (THz) wave detection, acting as a hot electron bolometric mixer (g-HEB) in heterodyne detection~\cite{Lara-Avila1}. For g-HEBs, understanding the energy relaxation processes in the material is crucial as it directly impacts the device design, and is paramount to achieve high sensitivities and large device bandwidths desired in astronomical observations ~\cite{Klapwijk}. In general, decreasing the thermal relaxation rate and heat capacity improves the sensitivity of bolometers and calorimeters~\cite{roope,JP1}; this observation has triggered a number of studies on electron-phonon heat transport in graphene at sub-kelvin temperatures~\cite{efetov, FG, Finkelstein, XuDu, Schwab}. For epigraphene, little is known about relaxation processes, particularly when the material is doped close to the Dirac point and in the millikelvin temperature range, conditions at which g-HEBs  are expected to perform better. Previous studies of the energy relaxation mechanisms of epigraphene have been mostly limited to samples at high carrier densities and at liquid helium temperatures ~\cite{Baker, Huang}, or  on micron-sized devices where thermalization of hot carriers occurred via the metallic contacts  (i.e. diffusion cooling) ~\cite{Lara-Avila1}. 

\begin{figure}[ht]
	\centering
	\includegraphics [width=\columnwidth] {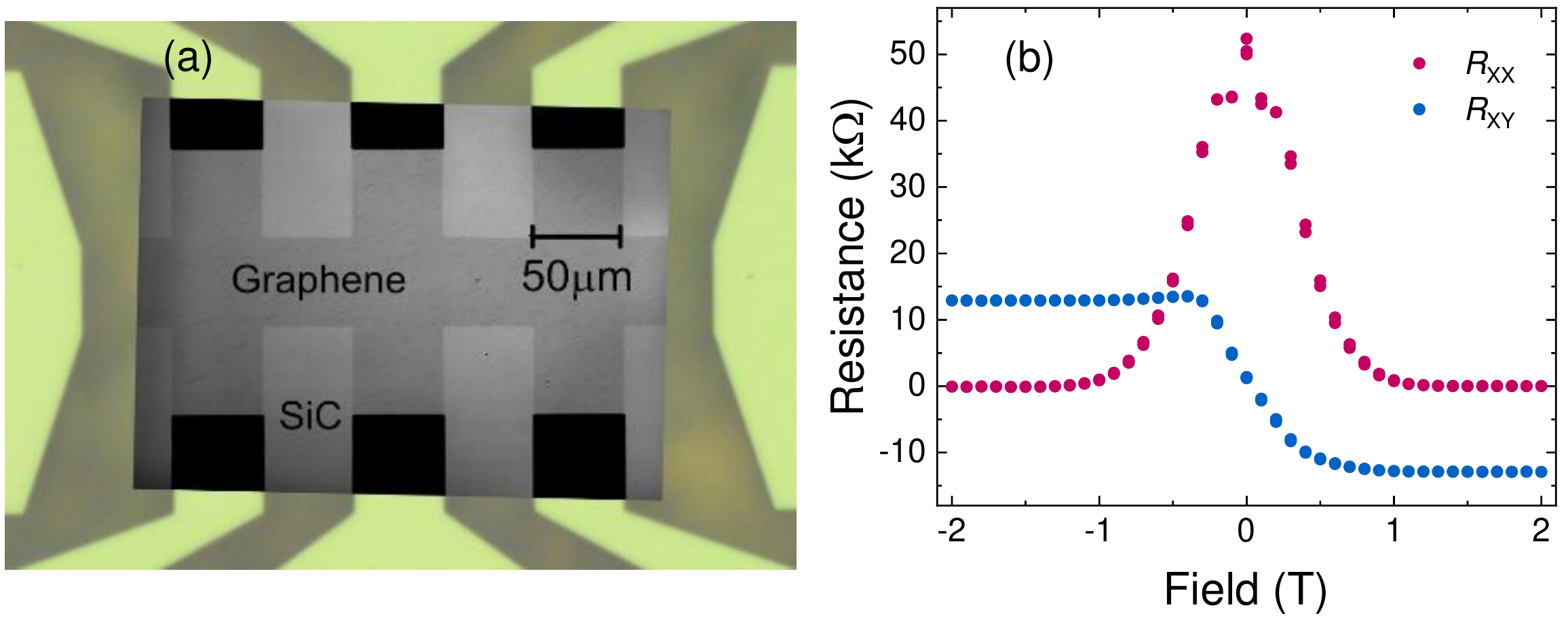}
	\caption{Hall bar device with doped epigraphene. (a) Optical micrograph showing the Hall bar device ($L=250$ $\mu$m, $w=50$ $\mu$m). The center overlay shows an image taken in transmission mode, which enables visualizing the epigraphene. (b) Example of DC electrical characterization of epigraphene Hall bar with hole type carrier density $p=1.7\times10^{10}$~cm$^2$, measured using $I=100$~nA and $T_0=2$~K. The sample demonstrates fully developed quantum Hall effect with $R_{XX}=0$ and $R_{XY}=h/2e^2$, the hallmark of monolayer graphene in magnetotransport.
		\label{fig1}}
\end{figure}
Here we present a study of energy relaxation in charge neutral epigraphene devices fitted with superconducting contacts, which act as thermal barriers that prevent heat leak from the contacts, thus enabling the study of energy relaxation processes in the graphene-silicon carbide system. We use in-situ thermometry down to sub-$100$~mK temperatures by measuring the sheet conductance of epigraphene Hall bar devices as they are locally heated by injecting current through Hall probes. This method provides a built-in thermometer, and its made possible thanks to the strong temperature dependence of resistance in lowly-doped epigraphene ~\cite{He2,Lara-Avila1}.

Epigraphene was grown on a 4H-SiC chips ($7\times7$~mm$^2$), which were encased in a graphite crucible and heated using RF heating to around \SI{1850}{\celsius} in an inert argon atmosphere of 1 bar~\cite{Virojanadara}. Transmission mode microscopy was used to ensure that the samples had high ($>90$~\%) monolayer coverage~\cite{Yager}. Device fabrication utilized standard electron beam lithography techniques, described in detail elsewhere~\cite{Yager2}. In short, epigraphene was patterned into Hall bar structure by oxygen plasma etching and superconducting metallic contacts were prepared with $30$~nm-thick aluminium contacts using a $6$~nm-thick adhesion layer of titanium. This is a proven technique to make transparent electrical contacts to graphene~\cite{Heersche, Mizuno}. The metallic layers were deposited using physical vapour deposition using electron beam evaporation. The finished device was spin-coated with molecular dopants and the final carrier density was tuned close to charge neutrality by annealing at $T =\SI{160}{\celsius}$~\cite{He2}. In order to test the device quality, initial DC electrical characterization was performed using quantum Hall measurements in a PPMS (Physical Property Measurement System from Quantum design) liquid helium cryostat ($2-300$~K) with a superconducting magnet providing fields up to $14$~T. Sub-kelvin measurements were performed in a dilution refrigerator.


Figure~\ref{fig1}(a) shows an optical micrograph of the doped epigraphene Hall bar used for our study, with channel length $L=250$ $\mu$m and width $w=50$ $\mu$m. Quantum Hall measurements at $2$~K were used to verify the quality of the devices (Fig.~\ref{fig1}(b)). The sample shows fully developed quantum Hall effect, with vanishing longitudinal resistance $R_{XX}=0$ and quantized transverse resistance $R_{XY}=h/2e^2$. This proves that the sample is of high quality monolayer epigraphene with spatially homogenous doping~\cite{He2,Cho, Tiwari}. The device is p-doped, with carrier density of $p=1.7\times10^{10}$~cm$^{-2}$, and mobility  $\mu=14,500$~cm$^2/$Vs. 
\begin{figure}[ht]
	\centering
	\includegraphics [width=\columnwidth] {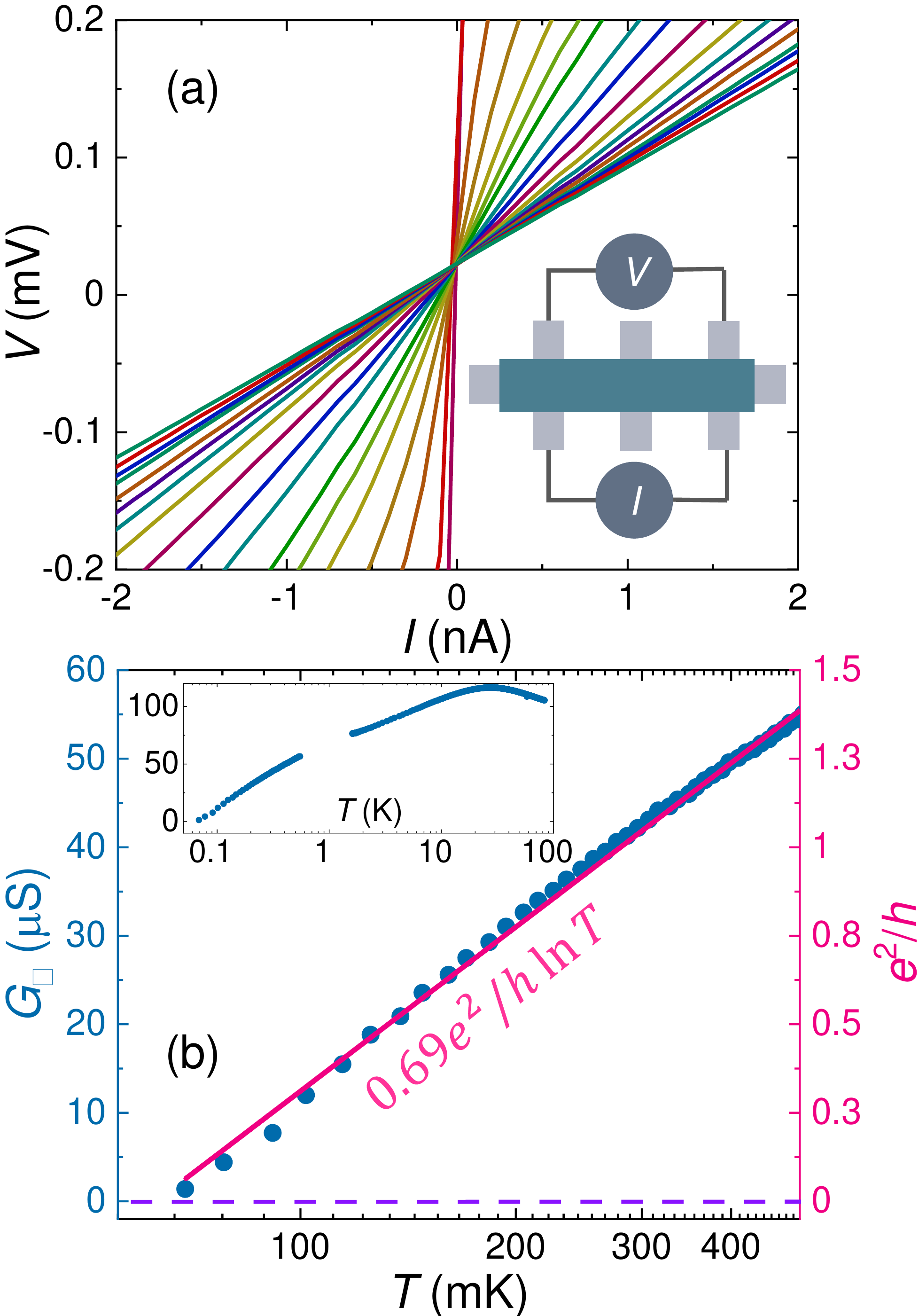}
	\caption{Low temperature electrical transport measurement of epigraphene. (a) Current-voltage characteristics ($IV$) measured in a four-probe configuration (inset) across the entire graphene channel. The different curves refer to different temperatures from about 50 mK up to 500 mK. (b) Zero bias differential sheet conductance on the logarithmic temperature scale in the same range as in (a). The inset displays the same measurements in a wider range of temperature. The pink line shows the logarithmic fit to the experimental data.
		\label{figA}}
\end{figure}

In Fig.~\ref{figA}(a) we show the current-voltage ($IV$) characteristics measured in a dilution refrigerator at various temperatures ranging from about $50$~mK up to $500$~mK. The four-probe configuration for these measurements is shown in the inset of Fig.~\ref{figA}(a). The temperature dependence of zero-bias differential sheet conductance of the device is seen in Fig.~\ref{figA}(b), extracted from the average slope of the $IV$ curve  over a current range of a few pA around zero current. The vertical scale on the right (shown in blue color) is the sheet conductance expressed in units of  the conductance quantum $\sigma_0=e^2/h$. The  inset of Fig.~\ref{figA}(b) shows the temperature dependence of the sheet conductance in a wider temperature range. Data in the two separate ranges of temperature were measured using different set-ups. We observe approximately logarithmic-in-temperature dependence of the sheet conductance of graphene, $\sigma (T)=\sigma_1+A\sigma_0 \ln(T)$, with $\sigma_0 = e^2/h \approx3.9 \times 10^{-5}$ S, where $e$ is the elementary charge, and $h$ is the Planck's constant. The slope of the logarithmic term quantifies the strength of the quantum corrections in the material, and for this p-doped sample, $A\approx0.69$, higher than $A\approx0.30$ reported for n-doped samples ~\cite{Lara-Avila1}. Studies in magnetic field would be required to verify if a higher $A$ is the result of enhanced electron-electron interactions or quantum interference effects in p-doped samples.

For thermal characterization of the device, we have considered four contributions when analysing the local thermal balance of the epigraphene structure. These are (i) the thermal conductance from epigraphene to the phonon bath $G_{\rm th}$, (ii) the lateral thermal conductance along the epigraphene sheet $\kappa$, (iii) the thermal conductance of the substrate partly shunting thermally the epigraphene ($G_{\rm SiC}$), and finally (iv) the thermal conductance to the superconducting leads ($G_{\rm out}$) to which the Hall bar structure is connected. For an ideal measurement of epigraphene properties only, the two first ones ought to dominate and the two others should not contribute to the heat currents. Yet, as we argue below, our measurements, together with  estimates of substrate material properties, and assuming that the Wiedemann-Franz law is approximately valid for epigraphene, the shunting effect (iii) exceeds the thermal conductance along the epigraphene sheet by several orders of magnitude. 

\begin{figure}[ht]
	\centering
	\includegraphics [width=\columnwidth] {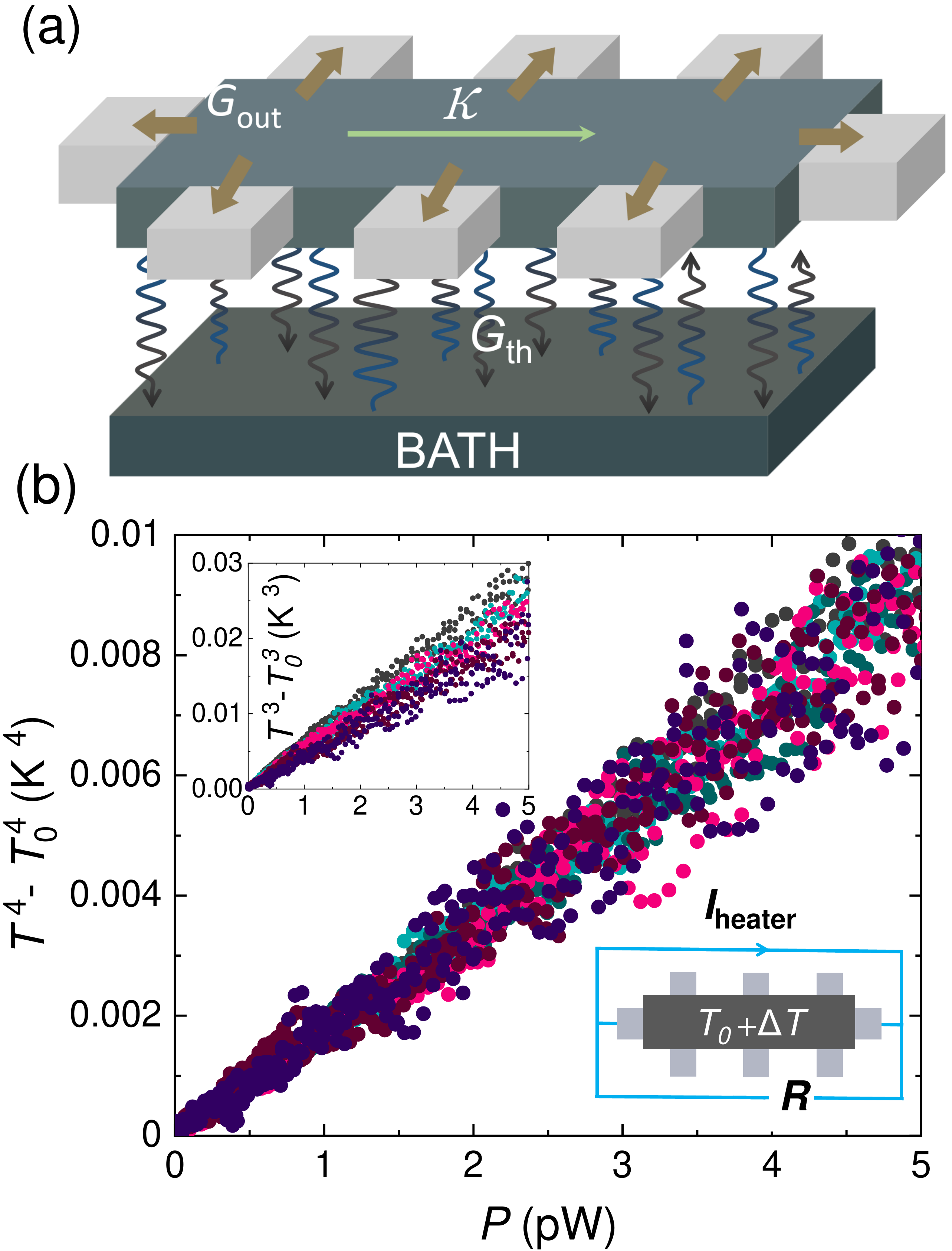}
	\caption{Electron phonon coupling in epigraphene doped close to Dirac point. (a) The dominant heat transport component is the conductance to the phonon bath, $G_{\rm th}$, and we can neglect the lateral conductance $\kappa$ and the leak into the leads $G_{\rm out}$. (b) The power $P=IV$ dependence of the difference in the fourth power of the temperatures of the epigraphene electrons and the substrate, $T_e$ and $T_0$, respectively. The substrate temperatures range is from 170-306 mKfor each data sets shown in different colours. The setup for this measurement is schematically presented in the lower inset in (b). The heat flow obeys the law $P=\Sigma_4 A (T_e^4-T_0^4)$ for graphene in the clean limit. The upper inset shows the corresponding plot for graphene in the dirty limit, with a power dependence with 3rd power in temperature.
		\label{figB}}
\end{figure}

In the measurements of the hot electron effect in the epigraphene device,  depicted in Fig. \ref{figB}(a), the Joule power $P=IV$ generated by the bias current leads to an increase of the electronic temperature of graphene, which is measured through monitoring the sheet conductance of the material. With superconducting contacts acting as heat barriers, the heat flow through the contacts is negligible, i.e. $G_{out} \approx 0$ and the only energy relaxation pathway in the system is through the phonon bath [i.e. item (i)] . Figure~\ref{figB}(b) is a collection of such measurements in the temperature range 170 mK to 306 mK, showing that the heat flux $P$ from the electron system at temperature $T_e$ to the phonon bath (at constant bath temperature $T_0$) follows the law 
\begin{equation}\label{P_eph}
P_{\rm ep}=\Sigma_n A (T_e^n-T_0^n)
\end{equation}
with $n=4$, and  $\Sigma_4\approx 0.04$ WK$^{-4}$m$^{-2}$. This power law is consistent with theoretical predictions for graphene in the clean limit ~\cite{FG, XuDu}, and with previous reports in highly n-doped epigraphene ($n = 1.63 \times10^{12}$ cm$^{-2}$ ) and at temperatures up to $T = 10$ K ~\cite{Baker,Huang}. For comparison, the inset of Fig.~\ref{figB}(b) shows the $T^3$ dependence of electron-phonon heat current for graphene in the dirty-limit. We conclude that  $n=4$ describes the experiment more closely, with data at various temperatures collapsing better on the same line. 

The electrical conductance across the Hall bar, e.g. between two adjacent Hall probes with a distance of $\sim 50~\mu$m, is $G\sim 10^{-5}~\Omega^{-1}$ (see Fig. \ref{figA}b). Using Wiedemann-Franz law we obtain $\kappa=G\mathcal{L}_0T\simeq 4\times 10^{-14}$~W/K for the thermal conductance in this geometry at $T=0.2$~K. Here $\mathcal{L}_0=2.4\times 10^{-8}~{\rm W}\Omega{\rm K}^{-2}$ is the Lorenz number. It is then evident that the electron-phonon conductance dominates and shunts the system thermally as under the same conditions $G_{\rm th}\equiv dP_{\rm ep}/dT_e= 4\Sigma_4 A_\square T_e^3\sim 3\times 10^{-12}$ W/K for the area under the heater. Since the thermal conductance of the 3D structure is large, we conclude that the role of $\kappa$ is negligible in our measurements.


The thermal characterization allows to calculate the noise-equivalent power (NEP), an important figure of merit of a bolometer. In the limit where this figure is determined by fundamental energy fluctuations, it assumes value ${\rm NEP}=2\sqrt{G_{\rm th}k_{\rm B}T^2}$. In our case $G_{\rm th}=4\Sigma_4AT^3$ from the electron-phonon measurement, $T=170$~mK, and $A=4\times 45\times 45\times 10^{-12}$~m$^2$ we obtain ${\rm NEP}\sim 3\times 10^{-18}~{\rm W}/\sqrt{{\rm Hz}}$. This is still not record-low because of the large area in our sample and relatively high $T$. In order to improve the NEP, one needs to reduce the area of the detector and operate it at lower $T$. One can realistically make $A\sim 10^{-11}$~m$^2$ and operate at about 10~mK. In this case the projected ${\rm NEP}\sim 10^{-22}~{\rm W}/\sqrt{{\rm Hz}}$, which would outperform the current experimental state-of-the-art~\cite{roope}. Although here we do not present heat transport data below $170$~mK, we believe that lower $T$ measurements are feasible: applying a moderate magnetic field of $\sim 50$~mT perpendicular to graphene sheet restores the conductance of it down to $T=10$~mK.

In summary, we have demonstrated sensitive in-situ thermometry by measuring the sheet conductance of the epigraphene sheet down to sub-$100$~mK temperatures, providing a way for  sensitive calorimetry with a built-in thermometer. The coupling of the epigraphene electrons to the phonon bath dies off more slowly with decreasing temperature than in metal films $(T^3...T^4$ versus $T^4...T^6)$~\cite{JPrmp}; therefore the advantage in operating at the very low temperatures is not quite that obvious in case of epigraphene as compared to metals, where furthermore the proximity superconductivity can be used for enhancement of sensitivity. Yet, the extremely small heat capacity of the epigraphene sheets at low temperatures leads to very fast thermal relaxation times of the order of $10$~ps, making, together with in-situ thermometry and weak $G_{\rm th}$, epigraphene bolometer an attractive choice for terahertz applications. 

\begin{acknowledgements}
This work was jointly supported by the Swedish Foundation for Strategic Research (SSF) (Nos. GMT14-0077 and RMA15-0024), Chalmers Excellence Initiative Nano, and European Union’s Horizon 2020 research and innovation programme under Marie Sklodowska-Curie Grant Agreement No 766025. This work was performed in part at Myfab Chalmers. We acknowledge the facilities and technical support of Otaniemi research infrastructure for Micro and Nanotechnologies (OtaNano). We thank the Russian Science Foundation (Grant No. 20-62-46026) for supporting the work.
\end{acknowledgements}

\end{document}